\documentstyle[prl,psfig,aps]{revtex}

\begin{document}
\draft

%
%

\title{Multi-phonon Resonant Raman Scattering Predicted
in LaMnO$_3$ from the Franck-Condon Process {\it via} Self-Trapped Excitons} 

\author{Vasili Perebeinos and Philip B. Allen}
\address{Department of Physics and Astronomy, State University of New York,
Stony Brook, NY 11794-3800}
\date{\today}
\maketitle

\begin{abstract}
Resonant behavior of the Raman process is predicted when the laser 
frequency is close to the orbital excitation energy of LaMnO$_3$ at 2 eV. 
The incident photon creates a vibrationally excited 
self-trapped ``orbiton'' state from the 
orbitally-ordered Jahn-Teller (JT) ground state.
Trapping occurs by local oxygen rearrangement.  Then the Franck-Condon
mechanism activates multiphonon Raman scattering.
The amplitude of the $n$-phonon process is first order in the  
electron-phonon coupling $g$.  The resonance occurs {\it via}
a dipole forbidden $d$ to $d$ transition.  We previously suggested that
this transition (also seen in optical reflectivity) becomes allowed because of
asymmetric oxygen fluctuations.  Here we calculate the magnitude of the 
corresponding matrix element using local spin-density functional 
theory.  This calculation agrees to better than a factor of two with
our previous value extracted from experiment.  This allows us 
to calculate the absolute value of the Raman tensor for multiphonon 
scattering.  Observation of this effect would be a direct confirmation
of the importance of the JT electron-phonon term and the presence of
self-trapped orbital excitons, or ``orbitons.''
\end{abstract}


%
%

\section{introduction}

Manganese oxide materials attract 
attention because of the ``colossal magnetoresistance''
(CMR) phenomenon \cite{CMR}, and because of a
very rich phase diagram \cite{phase} of ground states with competing
order parameters. 
The Mn$^{+3}$ ion of the parent LaMnO$_3$ compound has 
$d^4$ $(t_{2g}^3,e_g^1)$ configuration with an inert $t_{2g}$ core (spin 3/2).
The half-filled doubly degenerate $e_g$ orbitals 
($d_{x^2-y^2}$, $d_{3z^2-r^2}$) 
are Jahn-Teller (JT) unstable.  A symmetry-breaking oxygen distortion 
(resulting in the Mn-O bond lengths of 1.91, 2.18 and 1.97 \AA 
\cite{Rodriguez}) lowers the energy of the occupied orbital. 
The corresponding orbitally ordered state sets in at $T_{\rm JT}=750$ K
with $x-$ and $y-$oriented $e_g$ orbitals in the $x-y$ plane with 
wavevector $\vec{Q}=(\pi, \pi, 0)$ \cite{Murakami}.  The orbital order drives 
antiferromagnetic ($A$-type) spin order \cite{Moussa} below the Neel 
temperature $T_N=140$ K.

There is still controversy about the origin of the orbitally ordered state.
Strong electron-electron correlations may lead to orbital order {\it via}
the superexchange interaction \cite{Kugel} which lifts degeneracy of the $e_g$ 
states \cite{Tokura}.  In another scenario proposed by Millis \cite{Millis1} 
the Jahn-Teller electron-phonon (e-ph) interaction $g$ \cite{Kanamori} causes 
the orbital order and contributes to CMR. Extensive numerical work by 
Dagotto {\sl et al.} \cite{Hotta} showed that the two approaches give
qualitatively similar answers.  We prefer a model where the
JT interaction $g$ plus large Hubbard $U$ and Hund energy $J_H$ leads to single 
occupancy of the Mn $e_g$ levels and a gap to on-site $d$ to $d$
excitations, rather than assigning the gap purely to Coulomb interactions
as in a multi-orbital Hubbard model.  We believe that the importance of the
JT interaction is evident at low hole doping 
of La$_{1-x}$Sr(Ca)$_x$MnO$_3$, whose insulating nature is naturally 
explained by formation of the anti-Jahn-Teller polarons \cite{Allen1}.

\begin{figure}
\psfig{figure=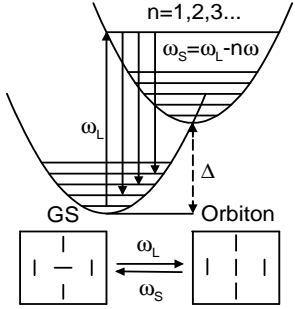,height=2.5in,width=3.0in,angle=0}
\caption{Schematic Franck-Condon mechanism for the multiphonon Raman process 
via the orbiton intermediate state. 
The lowest energy configuration of the orbiton has 
energy $\Delta$ and large oxygen distortions from the JT ground state (GS).
The most probable intermediate state (the strongest resonance of the Raman 
process) occurs at
$\omega_{\rm L}\approx 2\Delta\approx$ 2eV, rather than $\omega_L=\Delta$.}
\label{fig1}
\end{figure}

In this paper we present a detailed prediction of resonant 
multiphonon Raman features, whose observation would be a direct measure
of the importance of the JT electron-phonon term.
When the oscillator potential curves of ground and excited states are 
displaced relative to each other, then vibrational Raman scattering 
is activated by a Franck-Condon (FC) two-step mechanism.
Our Hamiltonian for LaMnO$_3$, with $U\rightarrow\infty$, leads to
a picture where the ground state and low-lying excited states
are simple products of localized orbitals, one per atom.
In the first step of the FC Raman process, the incident photon 
creates an orbital defect in the ordered JT ground state 
(one Mn ion has the upper rather than
the lower state of the JT doublet occupied.) 
This Frenkel exciton (also called an ``orbiton'') is self-trapped
\cite{Allen2} by oxygen rearrangement from the JT state. 
The FC principle has the oxygen positions undistorted
during optical excitation, producing a vibrationally excited state
of the orbiton.  In the second step of the Raman process, this
virtual excitation decays back to the orbital ground state, but not
necessarily the vibrational ground state.  The amplitude for
ending in a vibrationally excited state is determined by 
displaced-oscillator overlap integrals.  This
allows $n$-phonon resonant Raman scattering with amplitude proportional to 
the first order of the e-ph interaction \cite{Shorygin}. 
The process is illustrated in Fig. \ref{fig1}.
In the conventional Raman scattering process, where electronically excited 
states do not alter atomic positions, the amplitude of the $n$-phonon peak is 
proportional to the $n$th order of the e-ph interaction which is smaller by 
$n-1$ orders of magnitude.  The conventional 
process can be divided into three steps.
(1) The incident photon creates an electron-hole pair (or exciton).
(2) This electron-hole pair is scattered into another state by sequential
emission of $n$ phonons {\it via} $n$ powers of the e-ph 
interaction ${\cal H}_{\rm e-ph} \propto g$.  Higher-order interactions,
such as the electron-(two phonon) interaction also enter, but do not
increase the order of magnitude of the process.
(3) The electron-hole pair recombines, emitting a scattered photon.
In this formulation, the intensity of the two-phonon Raman process 
is smaller than one-phonon by several orders of magnitude 
10$^{-2}$-10$^{-3}$, determined by the $2n$-th power
of the ratio of e-ph to electronic energies.

\section{Franck-Condon mechanism}

We use a model Hamiltonian \cite{Allen1}, essentially the 
same as used by Millis \cite{Millis2}, with two $e_g$ orbitals per Mn atom, 
fully respecting the symmetries of the orbitals and the crystal.
The electron-phonon term ${\cal H}_{\rm JT}$ stabilizes the orbitally-ordered
ground state {\it via} a cooperative JT distortion.
Oxygen displacements along Mn-O-Mn bonds are modeled by 
local Einstein oscillators:
\begin{eqnarray}
{\cal H}_{\rm JT}&=&-g\sum_{\ell,\alpha} \hat{n}_{\ell,\alpha}
(u_{\ell,\alpha}-u_{\ell,-\alpha})
\nonumber\\
{\cal H}_{\rm L}&=&\sum_{\ell,\alpha}(P^2_{\ell,\alpha}/2M
+Ku^2_{\ell,\alpha}/2).
\label{hmodel}
\end{eqnarray}
The interaction ${\cal H}_{\rm JT}$ consists of linear energy
reduction of an occupied $d_{3x^2-r^2}$ orbital
(the corresponding creation operator is $c^{\dagger}_{x}$ and number
operator is $\hat{n}_{\ell,x}=c^{\dagger}_{x}(\ell)c_{x}(\ell)$)
if the two oxygens in the $\pm \hat{x}$ direction
expand outwards.   Similar terms are included for $\hat{y}$ and $\hat{z}$
oxygens if $d_{3y^2-r^2}$ or $d_{3z^2-r^2}$ orbitals are occupied.
The strength $g$=1.84 eV/$\AA$ of the JT coupling $g$ determines 
the JT splitting of the orbitals $2\Delta=1.9$ eV, and was chosen to agree 
with the lowest optical conductivity peak \cite{Jung}.
The displacement $u_{\ell,\alpha}$ is measured from the cubic
perovskite position of the nearest oxygen in the $\hat{\alpha}$-direction
to the Mn atom at $\ell$.  The oxygen vibrational energy  
$\hbar\omega=\hbar\sqrt{K/M}=0.075$ eV is taken
from a Raman experiment \cite{Raman}.
In addition there is an on site Coulomb repulsion U and a large Hund 
energy $J_H$. In the limit $U\rightarrow\infty$ and $J_H\rightarrow\infty$,
electronic motion at  half-filling is suppressed due to single  
occupancy of the Mn sites; additional orbital splitting caused by
superexchange interactions is left out.

The Hamiltonian ${\cal H}={\cal H}_{\rm JT}+{\cal H}_{\rm L}$
gives an orbitally ordered ground state:
\begin{equation}
|0,0>=\prod_{\ell}^A c_X^{\dagger}(\ell)
    \prod_{\ell^{\prime}}^B c_Y^{\dagger}(\ell^{\prime})|\{0\}>,
\label{gs1}
\end{equation}
where $\{0\}$ refers to the lattice vibrational ground state with oxygen
atoms in distorted equilibrium positions (Van Vleck $Q_2$-type distortions)
$u_{\ell\pm x}=\pm u_0$, $u_{\ell\pm y}=\mp u_0$, $u_{\ell\pm z}=0$ 
if $\ell \in A$ sublattice ($\exp(i\vec{Q}\cdot\vec{\ell})=1$)
and opposite sign distortions 
if $\ell \in B$ sublattice ($\exp(i\vec{Q}\cdot\vec{\ell})=-1$.)  
The magnitude of the 
distortion $2u_0=\sqrt{2\Delta/M\omega^2}$=0.296 {\AA} agrees
with neutron diffraction data \cite{Rodriguez} within 10-15\%,
confirming the internal consistency of the model.
Operators $c_{X,Y}^{\dagger}$ create electrons with orbitals 
$\Psi_{X,Y}=-(d_{3z^2-r^2}\mp d_{x^2-y^2})/\sqrt{2}$ alternating
on  $A$ and $B$ sublattices.
The lowest-lying electronic excitation of the Hamiltonian (\ref{hmodel}) is a 
self-trapped exciton or orbiton \cite{Allen2}, which gives a broad line 
in the optical conductivity \cite{Jung} centered at $2\Delta\approx2$ eV.

In the $n$-phonon Raman process, incident light of frequency $\omega_L$ is 
scattered with a shifted frequency 
$\omega_S=\omega_L-n\omega$.  The Raman cross-section tensor  
${\sf R}^n$ can be found as following \cite{Cardona}:
\begin{eqnarray}
\frac{\partial^2 {\sf R}^n}{\partial \omega_R\partial\Omega}&=&
\frac{\sigma_0}{m_e^2}\frac{\omega_S^2}{\omega_L^2}
\left|\sum_{\{m\},i}\frac{<0,n|\hat{\varepsilon}_L\cdot\vec{p}|i,m>
<i,m|\hat{\varepsilon}_S\cdot\vec{p}|0,0>}
{\Delta+N\{m\}\hbar\omega-\hbar\omega_L+i\gamma_m} + {\rm NRT}\right|^2 
\delta(\omega_R-n\omega)
\label{phram}
\end{eqnarray}
where $\sigma_0=r_e^2$ is a Compton cross section 
($r_e=e^2/m_ec^2$).
The summation goes over all electronic states $i$ and all the 
corresponding vibrational quanta $\{m\}$. 
$N\{m\}=m_{x}+m_{y}+m_{z}+m_{-x}+m_{-y}+m_{-z}$ is
the total number of vibrational quanta. 
The nonresonant term (NRT) is obtained 
from the resonant term by permuting $\hat{\varepsilon}_L$ with 
$\hat{\varepsilon}_S$ and changing $-\omega_L$ to $\omega_S$. The imaginary 
frequency $\gamma_m$ takes into account the finite lifetime of the 
intermediate state $|i,m>$.
The final state $|0,n>$ has an electronic ground state plus $n$ vibrational 
quanta. Summation over all the possible states with a total 
number of $n$ vibrations is assumed in Eq. (\ref{phram}). For example, one 
phonon can be excited on any of the 6 neighboring oxygen atoms; two phonons
can be excited in 21 ways; three phonons in 56 ways, 
and so on. The ground state couples to excited 
electronic states by the electron-radiation Hamiltonian
$(\vec{p}\cdot\vec{A})$.  In LaMnO$_3$, 
we consider only the lowest excited electronic state with an orbital flip
(e.g. $|X>$-type to $|Y>$-type \cite{Allen1}). By neglecting coupling to 
higher electronic states we underestimate (perhaps by a significant
factor) the first order Raman peak intensity.  
For multiphonon Raman scattering, in first 
approximation, we assume that only orbiton intermediate states contribute.

\section{LSDA Dipole Matrix elements}

To evaluate dipole matrix elements in Eq. (\ref{phram}) we use 
the FC approximation. The wavefunctions $|0,0>$, $|i,m>$ and 
$|0,n>$ are written as products of vibrational wavefunctions
$\chi(\vec{R})$ dependent on the oxygen positions $\vec{R}$, and 
electronic wavefunctions $\psi(\vec{r},\vec{R})$ dependent on 
both electronic $\vec{r}$ and vibrational coordinates. 
The electronic dipole matrix element is a $d$ to $d$ transition and therefore
forbidden when the surroundings are symmetric.  Searching for a mechanism
to activate this transition, we notice that
an asymmetric oxygen displacement will cause Mn $e_g$ orbitals to acquire an 
admixture of $4p$ character. A typical mixing coefficient is
\begin{eqnarray}
\gamma_z&=&\int d\vec{r} \psi_{3z^2-r^2}\frac{\partial V}{\partial u_z}
\psi_{z}/(\epsilon_d -\epsilon_p)
\label{gamma}
\end{eqnarray}
where $\psi_z$ is an orbital of $p$ character, and $\partial V/\partial u_z$ is 
the perturbation caused by a displacement of oxygen $\ell+\hat{z}$ 
in $\hat{z}$ direction. The corresponding allowed optical matrix element is
\begin{eqnarray}
d_z&=&\int d\vec{r} \psi_{3z^2-r^2} p_z \psi_{z}.
\label{dd}
\end{eqnarray}
The resulting dipole matrix element is 
\begin{eqnarray}
&&<i,m|\hat{\epsilon}_{S,L}\cdot\vec{p}|0,n>=
\sum_{\alpha=x,y,z}\gamma_{\alpha}d_{\alpha} \epsilon_{\alpha}
<m_{x} m_{-x} m_{y} m_{-y} m_{z} m_{-z}|
(u_{\alpha}+u_{-\alpha})
|n_{x} n_{-x} n_{y} n_{-y} n_{z} n_{-z} >
\label{modelme}
\end{eqnarray}

If the ground state is described by Eq. (\ref{gs1}), then from symmetry 
one can show that $\gamma_xd_x=\gamma_yd_y=-\gamma_zd_z/2$. In our 
previous work \cite{Allen2} a 
phenomenological parameter $\gamma d=-\gamma_zd_z$ 
was introduced to account for the observed spectral weight of the optical
conductivity peak due to the self-trapped exciton. 
The oscillator strength $f$ defined as
$\int d\omega\sigma(\omega)=(\pi Ne^2/2m_e\Omega)f$ is equal in our model to
$f_{zz}=2((\gamma d)^2/m_eM\omega^2)/(2\Delta/\hbar\omega+1)$,
and $f_{xx}=f_{yy}=f_{zz}/4$. 
Here $N/\Omega$ is the Mn atom concentration.
The measured spectral weight $540 \Omega^{-1}{\rm cm}^{-1}$ eV
of the lowest broad line centered at 2 eV \cite{Jung}
corresponds to $f_{\rm exp}=0.113$ or $\gamma d=1.7$.
Here we use density functional theory (DFT) to calculate  an induced 
dipole matrix elements to test whether our choice of the parameter $\gamma d$ 
from the optical data was justified.

LaMnO$_3$ has been extensively studied by first-principles approaches 
\cite{Pickett} including the local-spin-density approximation (LSDA) of DFT, 
LDA+U, and Hartree-Fock methods.  Information about 
electronic and magnetic structure, 
and about electron-phonon and Coulomb interactions has been 
obtained.  Here we use LSDA to calculate the
dipole matrix element for $d$ to $d$ transitions as it is
induced by asymmetric oxygen distortions.  Rather than calculating
$\partial V/\partial u_z$ and doing perturbation theory as in Eqs.
(\ref{gamma},\ref{dd}), we directly calculate $<i|\epsilon\cdot p|0>$
in the presence of an imposed asymmetric oxygen distortion.

To solve the LSDA equations we use the plane-wave pseudopotential method 
\cite{Chetty,detail} with a spin-dependent exchange-correlation potential 
\cite{Ceperley} and a 
supercell approach. Calculations were done for a 10-atom perovskite supercell 
with only $Q_2$-type JT oxygen distortions
(the rotation of the MnO$_6$ octahedra is omitted.) 
The point group symmetry of the cell is thus
$D_{4h}$.  For convenience, the magnetic order is 
taken to be ferromagnetic with 4.00 $\mu_B$ 
spin magnetization per formula unit.  The lattice constant 3.936 {\AA} 
gives the same cell volume as observed for LaMnO$_3$. The magnitude of the 
in-plane oxygen distortions along the Mn-O-Mn bonds is $u_0=0.14$ \AA. 
The self-consistent charge density was calculated 
using six special {\bf k} points 
in the irreducible wedge of the Brillouin zone.  Then the 
self-consistent potential was used to calculate wavefunctions at the 
$\Gamma$ point. 

\begin{figure}
\psfig{figure=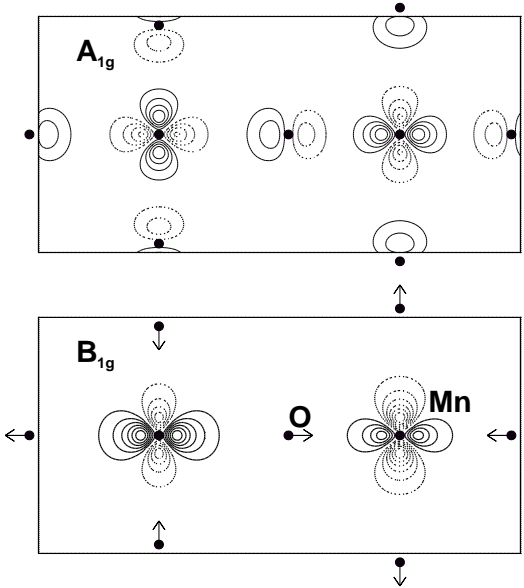,height=3.1in,width=4.5in,angle=0}
\caption{The LSDA wavefunctions $\Psi(\vec{r})$ at $\vec{k}=(0,0,0)$ point 
for the Mn $e_g$ states $B_{1g}$ (occupied) and $A_{1g}$ (empty).
Arrows indicate the direction of the $Q_2$-type oxygen displacements 
in the $xy$ plane. 
Negative (dashed) and positive (solid) contour values are in units of 
0.1 $e^{1/2}/(a. u.)^{3/2}$}
\label{fig2}
\end{figure}

A symmetry analysis of the pseudowavefunctions for ${\bf k}=(0,0,0)$ 
and plots of $|\psi|^2$ around the Mn site 
allow us to distinguish Mn $e_g$ states from other states. 
The $e_g$ ($X$- and $Y$-type) orbitals form states of $A_{1g}$ and $B_{1g}$ 
symmetries. Two of the four Mn $e_g$ states are shown on Fig. \ref{fig2}.
By introducing a small displacement of all the 
apical oxygens in $+\hat{z}$ direction along the Mn-O-Mn bonds, one induces
a $p_z$ dipole matrix element ($A_{1g}$ to $A_{1g}$ and
$B_{1g}$ to $B_{1g}$).  Similarly,
in-plane oxygen displacement in the
$+\hat{x}$ direction induces a $p_x$ matrix element. 
In table \ref{tab} we present LSDA results for these matrices.
The imposed displacement has lowered the $D_{4h}$ symmetry of the supercell,
permitting transitions between 
states below and above the Fermi level which were previously labeled as  
$A_{1g}$ and $B_{1g}$.
The energies of the Mn $e_g$ states do not alter much (less than 0.04 eV) for 
small distortions (0.016 \AA) and the induced 
dipole matrix elements are linear with oxygen distortion.

It is convenient to measure the induced dipoles in units of 
$(m_eM\omega^2)^{1/2}=0.472  \hbar/(a_{\rm B})^2$, where $m_e$ and $M$ are 
electron and oxygen masses and $a_{\rm B}$ is the Bohr radius. 
The calculated dipoles $(\gamma d)_z=2.39$ and $(\gamma d)_{x,y}=1.9$ give 
the oscillator strengths $f_{zz}=0.43$ and  $f_{xx,(yy)}=0.27$.  These
answers are somewhat larger than experiment, but well within the expected
accuracy of our model.  This accuracy is limited by four factors:
(1) convergence of the LSDA result; (2) neglect of rotational distortions;
(3) simplification of magnetic structure to ferromagnetic; and
(4) applicability of LSDA to strongly-correlated electrons.
We tested the first by varying the number of plane waves used in the 
pseudopotential expansion.  The answers reported here used
a plane-wave cut-off $E_{\rm pw}=135$ Ry.  The value of
$\gamma d$ increases by 9\% at $E_{\rm pw}=120$ Ry and decreases
by 13\% at $E_{\rm pw}=100$ Ry, so the convergence error is estimated
at 10\%.  Sources (2) and (3) will causes errors of similar size, we
believe.  The biggest uncertainty is source (4).  In an unrelated
DFT study on the self-trapped-exciton in
crystalline NaCl \cite{Perebeinos}, we found evidence that LSDA 
incorrectly shifts the energy of localized states relative to
delocalized states.  However, the wavefunctions under consideration
here are quite well-localized and agree qualitatively with expectation.
Therefore we do not believe that point (4) is the source of a major error.
Therefore we believe that this calculation provides firm evidence 
that phonon-activation
is strong enough to account for the observation of $d$ to $d$
(orbiton) transitions in optical reflectivity, as we assumed in
our previous work, and as we now use to predict Raman spectra.

\section{The Raman tensor}

The absolute cross sections for the Raman process Eq. (\ref{phram}) can 
be evaluated using expression (\ref{modelme}):
\begin{eqnarray}
&&\frac{\partial^2 {\sf R}_{\alpha\beta}^n}{\partial \omega_R\partial\Omega}=
\sigma_0\frac{\omega_S^2}{\omega_L^2}
(\gamma_{\alpha}d_{\alpha}\gamma_{\beta}d_{\beta})^2
\delta(\omega_R-n\omega)
\sum_{\{f\}}\delta(n-N\{f\})
\left|\sum_{m=0}^{\infty}
\frac{\hbar\omega A_{\alpha\beta}(m,\{f\})}
{\Delta+m\hbar\omega-\hbar\omega_L+i\gamma_{m}} + {\rm NRT}\right|^2,
\label{absR}
\\
&&A_{\alpha\beta}(m,\{f\})=
\sum_{\{m'\}}\delta(m-N\{m'\})
<f|u_{\alpha}+u_{-\alpha}|m'><m'|u_{\beta}+u_{-\beta}|0>
\label{absA}
\end{eqnarray}
where the induced dipole matrix elements $\gamma d$ and displacements $u$ 
are measured in units of $(m_eM\omega^2)^{1/2}$ and 
$\sqrt{\hbar/M\omega}$ respectively. 
In order to evaluate vibrational overlap integrals 
$A_{\alpha,\beta}(m,\{f\})$ 
one needs the expressions for overlap integrals of
displaced harmonic oscillators,
\begin{eqnarray}
&&<n_1|n_2>=(-1)^{n_1-n_2}\sqrt{n_1!n_2!}\ e^{-\kappa^2/2}\kappa^{n_1-n_2}
\sum_{k=0}^{n_2}(-1)^k\frac{\kappa^{2k}}{k!(n_2-k)!(n_1-n_2+k)!},\ \ 
{\text if} \ \ n_1\geq n_2
\label{int1}
\\
&&<n_1|u|n_2>=\frac{\kappa}{\sqrt{2}}\left(1+\frac{n_2-n_1}{\kappa^2}\right)
<n_1|n_2>,
\label{int2}
\end{eqnarray}
where $\kappa$ is related to the Jahn-Teller gap as 
$\Delta=4\kappa^2\hbar\omega$. The overlap $<n_1|n_2>$ for $n_1<n_2$ 
has the same expression as (\ref{int1}) with  
$n_1$ and $n_2$ interchanged and the sign of the displacement 
$u_0=\sqrt{2}\kappa$ changed ($\kappa\rightarrow-\kappa$). 
When using expressions (\ref{int1}, \ref{int2}), the signs of the 
$\ell+\hat{x}$, $\ell-\hat{y}$ oxygens displacements are positive and 
$\ell-\hat{x}$, $\ell+\hat{y}$ are negative, if $\ell \in A$ 
(reverse signs if $\ell \in B$) and $\ell\pm\hat{z}$ oxygens are undisplaced. 
Evaluation of the overlap integrals $A_{\alpha\beta}(m,\{f\})$ is 
straightforward. For example, for the first order Raman peak, only four 
one-phonon final states will contribute:
\begin{equation}
A_{\alpha\beta}(m,\{f\}=1\ldots 4)=\delta_{\alpha,\beta}
\frac{e^{-\Delta}\Delta^{m}}{m!}\frac{m}{4\kappa\Delta}(\Delta+1-m)
\label{a1}
\end{equation}
with no contribution to the non-diagonal part of the tensor $\alpha\ne\beta$.
For second- and third-order Raman scattering, four and eight final states 
contribute to the non-diagonal part of the tensor:
\begin{eqnarray}
&&A_{xy(yz,zx)}(m,\{f\}=1\ldots 4)=
\pm\frac{e^{-\Delta}\Delta^{m}}{m!}\frac{m}{2\Delta},
\nonumber\\
&&A_{xy(yz,zx)}(m,\{f\}=1\ldots 8)=
\pm\frac{e^{-\Delta}\Delta^{m}}{m!}\frac{\kappa m}{\sqrt{2}\Delta^2}
(\Delta+1-m)
\label{a2}
\end{eqnarray}
The formulas for the diagonal parts of the higher-order Raman tensors are more 
complicated and will not be given here.

\begin{figure}
\psfig{figure=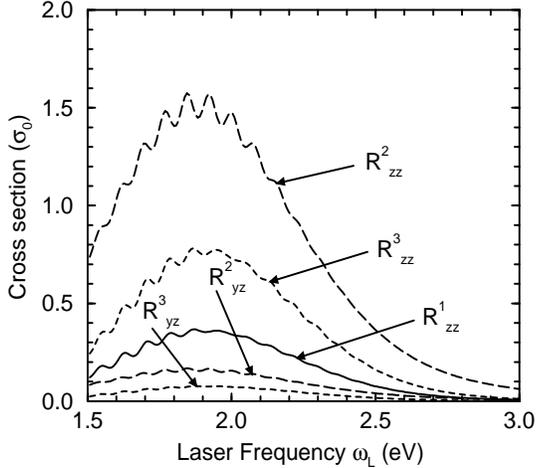,height=2.7in,width=3.0in,angle=0}
\caption{The absolute value of the multiphonon Raman cross section per 
solid angle {\sl versus} incident photon energy.  Resonant behavior is 
predicted for $\omega_L$ close to orbiton energy $2\Delta=1.9$ eV.
The damping constant is $\gamma_0=120$ cm$^{-1}$, 
the induced dipoles are 
$-\gamma_z d_z=2\gamma_x d_x=1.7$ $(m_eM\omega^2)^{1/2}$, cross section unit 
is $\sigma_0=(e^2/m_ec^2)^2$.
}
\label{fig3}
\end{figure}

To model the damping term 
$\gamma_m$ of vibrational level $m$,
we use expression $\gamma_m=\gamma_0\sqrt{m+1}$, 
as in a sequence of convolved Gaussians, intended to mimic
the local densities of phonon states on oxygen atoms. The value 
$\gamma_0=120$ cm$^{-1}$ was taken. 
The Raman cross section shown on Fig. 
\ref{fig3}  has a pronounced resonant behavior when the laser frequency 
$\omega_L$ approaches the orbiton energy $2\Delta$. 
The first-order cross-section 
as seen in Fig. \ref{fig3} is underestimated, because we  neglected
coupling to 
higher electronic levels in Eq. (\ref{absR}), whose contribution to 
multiphonon peaks is probably negligible. 
The polarization dependence of the cross-section is only due to the 
dipole matrix element effect, namely 
${\cal R}^{1,2,3}_{xx}={\cal R}^{1,2,3}_{yy}={\cal R}^{1,2,3}_{zz}/16$,
$R^{2,3}{yz}=R^{2,3}_{zx}=4R^{2,3}_{xy}$.
The anisotropy of the optical conductivity 
$\sigma_{xx}/\sigma_{zz}$ is quadratic and 
Raman intensity ${\cal R}_{xx}/{\cal R}_{zz}$
fourth-power in the dipole matrix-element anisotropy 
$\gamma_x d_x/\gamma_z d_z$. 
Actual occupied $e_g$ orbitals may be rotations of our idealized state  
Eq. (\ref{gs1}) in the $e_g$ space.
Even small deviations from Eq. (\ref{gs1})
might cause a noticeable change of the ratio $\gamma_x d_x/\gamma_z d_z$ 
from the value $0.5$.  Therefore the predicted anisotropies of 4 and 16 for 
optical and Raman spectra are not necessarily robust, but the relation 
${\cal R}_{xx}/{\cal R}_{zz}=(\sigma_{xx}/\sigma_{zz})^2$ should hold.

Published Raman measurements \cite{Raman} on undoped LaMnO$_3$ do not extend 
to the multiphonon region and resonant behavior has not been tested 
experimentally. Most experiments use the Ar$^{+}$ laser, for which
frequency  $\omega_L=2.41$ eV, we predict the multiphonon cross sections 
${\cal R}^2_{zz}=0.514$, 
${\cal R}^3_{zz}=0.243$, 
${\cal R}^2_{yz}=0.056$, 
${\cal R}^3_{yz}=0.025$ 
in units of $\sigma_0$ sr$^{-1}$.
Recently some features in Raman spectra on LaMnO$_3$ around 1100 cm$^{-1}$ 
were reported by several groups \cite{Bjornsson,Romero}. These are probably 
the effect we are predicting.

\section{Conclusion}

We advocate a picture of the orbitally ordered state
of LaMnO$_3$ where electron-phonon interactions (in a context of
large Hund and Hubbard energies) have a major influence.  Our picture
is disputed by other theorists \cite{Brink}.  Therefore we attempt here to
provide predictions which can qualitatively distinguish our
model from others.  In our model, the lowest electronic
excitation is the 1.9 eV transition across the Jahn-Teller gap,
modified by self-trapping to give a minimum gap half as large.
This has successfully described the observed \cite{Jung} optical
gap as a Franck-Condon broadened self-trapped exciton.
The present paper uses density-functional
theory to eliminate the need for a phenomenological
coupling $\gamma d$ to account for this transition.
  
As a more stringent test, we here predict a new feature
unique to the FC physics of the self-trapped exciton, namely
a sequence of resonant multiphonon Raman peaks. 
We predict the absolute values of the 
multiphonon Raman cross section tensors.

A hot luminescence process can also give rise 
to a multiphonon peaks. 
Incident light can excite an orbiton with a long lifetime which can recombine 
after vibrational energy $n\omega$ being lost in the intermediate state 
through anharmonic interaction. The question whether the Raman or 
hot luminescence mechanisms dominate the scattering intensity has an old 
history \cite{Lum}. In the Raman 
case the intensity of the higher order peaks should decrease and the width of 
the peak representing a convolved density of local phonon modes should 
increase. In the hot luminescence the intensities of the higher order peaks 
are of the same order and the width of the lines decrease with increasing 
order. In addition 
a strong emission peak can be observed at the exciton absorption edge. 
Raman techniques can serve as a direct probe of the orbiton excitation in 
LaMnO$_3$.

\acknowledgments
We thank D. Romero for encouragement on this problem, A. J. Millis for 
nudging us to do DFT calculations,
and M. Weinert for valuable discussions of the 
DFT results. 
This work was supported in part by NSF Grant No.\ DMR-0089492.

\begin{table}
\narrowtext
\caption{The absolute values of the induced dipole matrix elements 
$p_z$ (left) and $p_x$ (right) per unit oxygen displacements in 
units of $(m_eM\omega^2)^{1/2}$
between the LSDA wavefunctions of the $A_{1g}$ and $B_{1g}$ symmetry before an 
additional asymmetric oxygen distortion. The resulting oscillator strengths
for the transition between the occupied ($occ$) and empty ($em$)
states are $f_{zz}=0.43$ and  $f_{xx,(yy)}=0.27$, which corresponds to 
$(\gamma d)_z=2.39$ and $(\gamma d)_{x,y}=1.9$.}
\label{tab}
\begin{tabular}{c|cccc|cccc}
   & $A_{1g}^{occ}$ & $A_{1g}^{em}$ & $B_{1g}^{occ}$ & $B_{1g}^{em}$ & 
     $A_{1g}^{occ}$ & $A_{1g}^{em}$ & $B_{1g}^{occ}$ & $B_{1g}^{em}$  \\ 
\hline
$A_{1g}^{occ}$ & 0    & 3.37 & 0    & 0    & 0    & 1.83 & 2.03 & 0.90 \\
$A_{1g}^{em}$ & 3.37 & 0    & 0    & 0    & 1.83 &  0   & 1.55 & 0.05 \\
$B_{1g}^{occ}$ & 0    & 0    & 0    & 0.17 & 2.03 & 1.55 & 0    & 0.82 \\
$B_{1g}^{em}$ & 0    & 0    & 0.17 & 0    & 0.90 & 0.05 & 0.82 & 0    \\
\end{tabular}
\end{table}

\end{document}